\documentclass[aps,prb,twocolumn,pdfencoding=auto,superscriptaddress]{revtex4-2}
\usepackage[T1]{fontenc}
\usepackage[latin9]{inputenc}
\usepackage{amsmath,graphicx,amssymb}
\usepackage[pdftex, pdftitle={Article}, pdfauthor={Author},colorlinks,citecolor=blue,linkcolor=blue,urlcolor=blue]{hyperref}
\usepackage{amsfonts}
\usepackage{color}
\usepackage{cancel}

\usepackage{color}
\usepackage{soul}

\begin{document}

\title{Buckyballs directly probe inter-valley coherence in moire heterostructures}

\author{Tamaghna Hazra}
\affiliation{Institut f\"ur Theorie der Kondensierten Materie, Karlsruher Institut f\"ur Technologie, 76131 Karlsruhe, Germany.}
\affiliation{Institut f\"ur Quanten Materialien und Technologien, Karlsruher Institut f\"ur Technologie, 76131 Karlsruhe, Germany.}
%\affiliation{Center for Materials Theory, Rutgers University, Piscataway, New Jersey 08854, USA}
\thanks{tamaghna.hazra@kit.edu}

\vspace{10pt}

\begin{abstract}
Moire materials like twisted bilayer graphene have emerged as a rich playground of strongly correlated physics, where the effect of interactions can be drastically enhanced by tuning the non-interacting density of states via twist angle. 
A key feature of bilayer graphene at small twist angles is that the two valleys of graphene are effectively decoupled at the non-interacting level, leading to an emergent U(1) symmetry corresponding to the conservation of valley-charge. 
Theoretical studies invariably find that short-ranged interactions tend to break this symmetry, leading to inter-valley coherence (IVC). 
Here, I propose an experimental signature of IVC order by coupling with a closely related system where valley-mixing is guaranteed - the pentagonal disclinations of graphene. 
Such disclinations are naturally found in fullerenes. 
The tunneling current between twisted bilayer graphene and an appropriately oriented large buckyball has a sharp signature of the onset of IVC order. 
Successfully integrating these fullerenes in a scanning tunneling microscope would allow spatial resolution of the IVC order parameter on a sub-moire scale. 
Experimentally constraining the broken symmetries and order parameters in the multitude of emerging correlated insulators is a crucial first step to understanding the superconductivity that develops on doping these exotic `normal' states.
\end{abstract}
\maketitle

\section{Introduction}

The experimental demonstration of strongly correlation physics~\cite{li2010} culminating in the observation of correlated insulators and superconductivity~\cite{cao2018, cao2018a} in magic-angle twisted bilayer graphene (MATBG) triggered the explosion of interest in the field we now know as moire materials~\cite{andrei2021, he2021, mak2022}. 
The key insight from early continuum models was that the bandstructure of a twisted bilayer hosts extremely narrow isolated bands at a so-called magic angle~\cite{lopesdossantos2007, bistritzer2011, lopesdossantos2012}. 
This results in a tremendous enhancement in the non-interacting density of states, so that even the nominal electron-electron interactions in graphene would be strong-coupling in dimensionless units.
It was anticipated that some interesting electronic phase would emerge from the strongly correlated system at this fine-tuned magic angle. 
The sheer variety of novel phases that have emerged in MATBG and related systems~\cite{sharpe2019a, nuckolls2020, ledwith2020, park2021, serlin2020, xie2021a} since their experimental realization was perhaps harder to anticipate.

\begin{figure}
    \centering
    \includegraphics[width=0.5\linewidth]{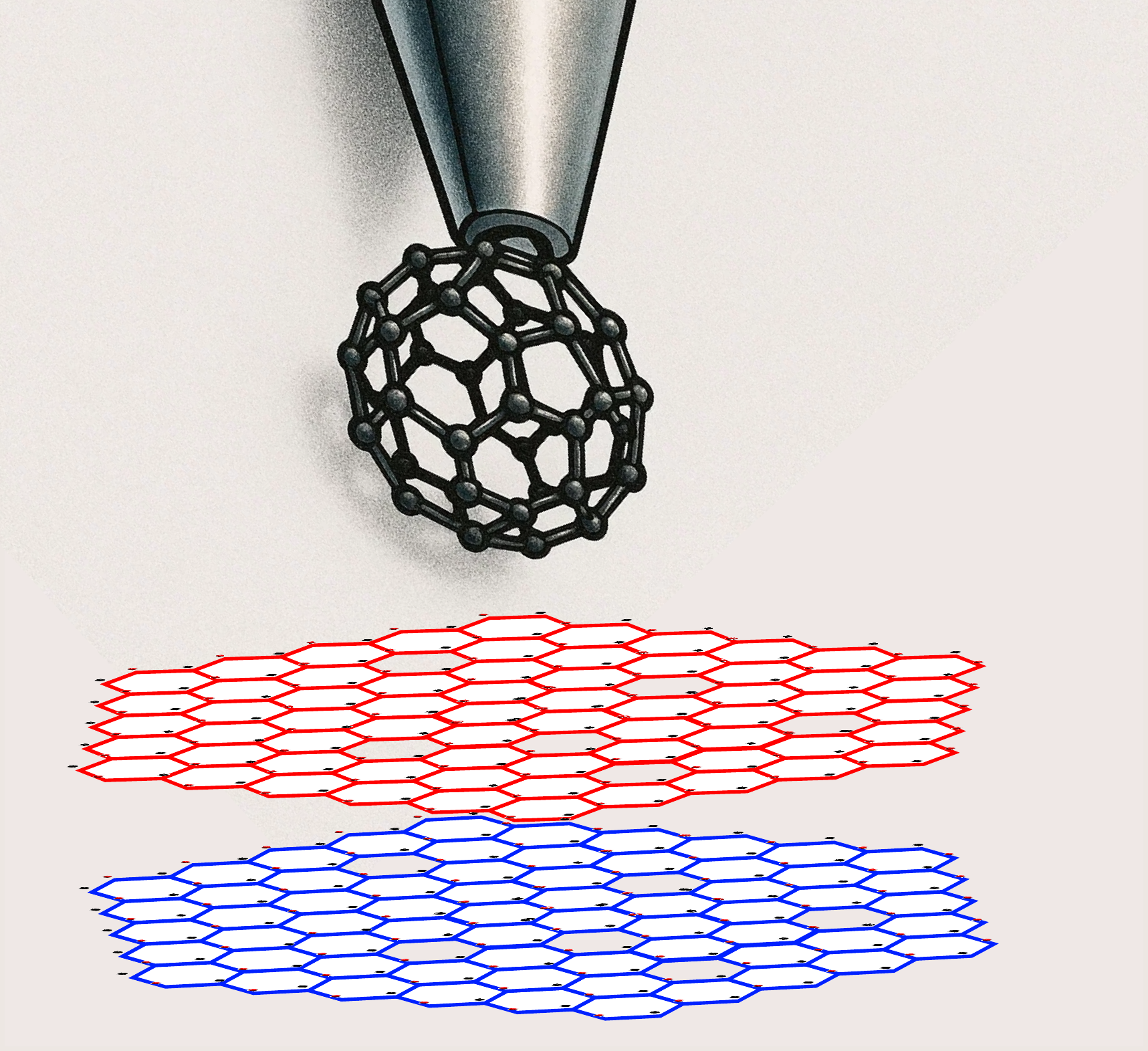}
    \caption{Tunneling geometry: A large buckyball suspended from metallic leads over a two-dimensional sample with putative inter-valley coherent order, say twisted bilayer graphene at the magic angle. For a sufficiently large buckyball suspended from a metallic lead, we approach the planar tunneling regime when the effective contact area is large compared with the moire unit cell.}
    \label{fig:geometry}
\end{figure}

\begin{figure*}[ht]
    \centering
    \includegraphics[width=0.7\textwidth]{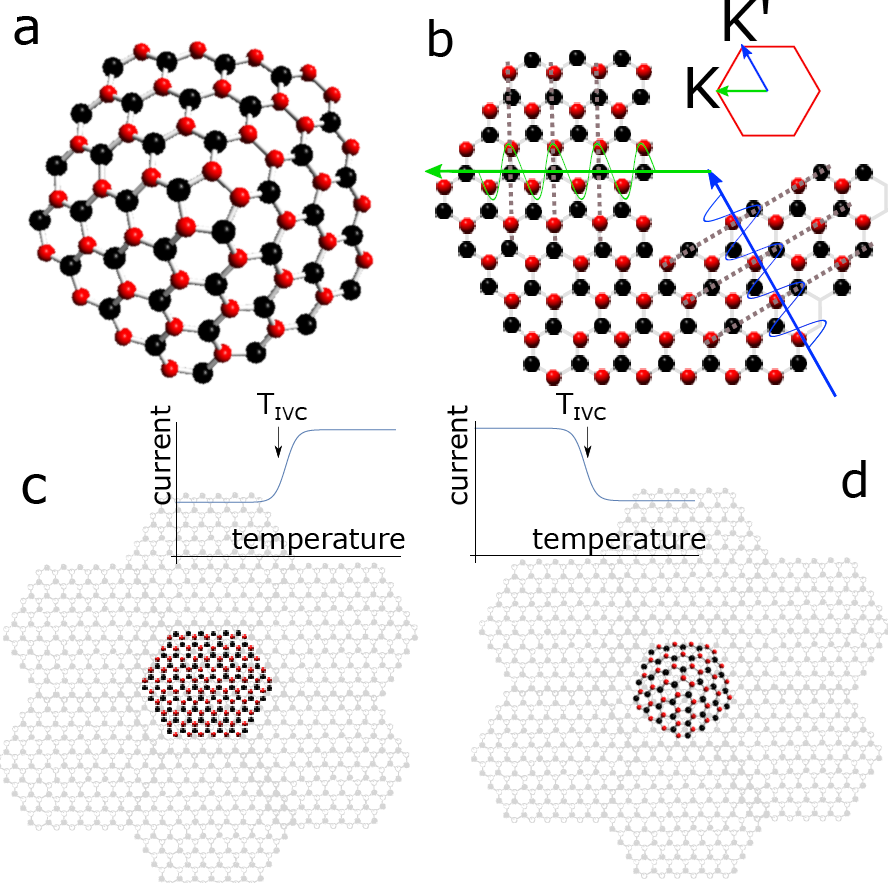}
    \caption{(a) Pentagonal disclination in graphene, also found at the vertices of fullerenes (b) Incoming wavepacket changes valley index at the disclination (c) Planar tunneling into MATBG from the hexagonal facets of the fullerene is suppressed at the onset of inter-valley coherence (d) Tunneling current into MATBG through the defect-bound states at the pentagonal disclination of the fullerene increases at the onset of inter-valley coherence. Note that the insets indicate the temperature dependence of the tunneling matrix element when the differential conductance $dI/dV$ is appropriately normalized by the tip and sample DOS. Inspired by \cite{cortijo2007}.}
    \label{fig}
\end{figure*}

The vast majority of putative phases have been traditional Landau symmetry-breaking ordered phases, see ~\cite{irkhin2018, kiese2022, zare2021, luo2022, nica2023} for examples of the complementary minority.  
In MATBG, in addition to the magnetic space group symmetries and $SU(2)$ spin-rotation, there is an additional ``approximate'' symmetry that emerges at low energies~\cite{guinea2018, kangSymmetryMaximallyLocalized2018, koshinoMaximallyLocalizedWannier2018, poOriginMottInsulating2018, xuTopologicalSuperconductivityTwisted2018, yuanModelMetalinsulatorTransition2018, zouBandStructureTwisted2018}. 
In essence, the interlayer hybridization modulates over the large moire lengthscale $a_M$, so that small-momentum ($\sim a_M^{-1}$) interlayer scattering of Dirac fermions effectively leaves the two valleys almost decoupled at small twist angles. This results in an approximate $U(1)$ symmetry corresponding to the conservation of valley quantum number of low-energy Dirac fermions. in the low-energy non-interacting description.

One of the common effects of adding interactions to this description is the spontaneous breaking of this $U(1)$ symmetry~\cite{guinea2018, kangSymmetryMaximallyLocalized2018, koshinoMaximallyLocalizedWannier2018, poOriginMottInsulating2018, xuTopologicalSuperconductivityTwisted2018, yuanModelMetalinsulatorTransition2018, zouBandStructureTwisted2018} - inter-valley coherence (IVC). This is not surprising since local interactions, unlike long-wavelength moir\'e potentials, have no reason to only allow small-momentum scattering. 
It is interesting to note that the particle-hole conjugate of the IVC order parameter $\langle c_{K+q,\alpha}^\dagger c_{K'+q,\alpha'} \rangle$ is essentially finite-momentum intra-valley pairing $\langle c_{K+q,\alpha}^\dagger \mathcal{T} c_{K-q,\alpha'}^\dagger \rangle$ where $\mathcal{T}$ is the time-reversal operator~\footnote{Particle-hole transformation historically has many definitions; here I define the particle-hole transform of a fermion bilinear as the action of charge conjugation and time-reversal on one fermion of the pair: $c_\alpha c_{\alpha'}\to c_\alpha \mathcal{T}\mathcal{C}c_{\alpha'}$. This definition is independent of any spatial symmetries, the presence of a lattice, with or without disorder or defects}. 
Both ordering tendencies are stabilized only above a finite threshold of interaction strength, being disfavored by the trigonal distortion of the Fermi surfaces centered on the $K$ and $K'$ valleys. 
In many strongly correlated superconductors, the fluctuations of the normal-state ordering drives pair formation and dictates pairing symmetry~\cite{scalapino2012}. 
It is possible that the proximate inter-valley coherence may play a similar role in driving intra-valley pairing in MATBG and related systems~\cite{murshed2022}. 
Intra-valley pairing has been shown to drive Kekule pair-density waves, chiral and helical topological superconductors, nematic and stripe superconductors in related models~\cite{lee2019, hsu2017, roy2010, tsuchiya2016}. Understanding IVC in moir\'e systems has important and interesting downstream consequences.

An important need of the hour is to directly and quantifiably probe the inter-valley coherence experimentally. 
Certain IVC states have an associated density wave order that can be readily observed with local probes such as scanning tunneling microscopy~\cite{nuckolls2023a,kim2023a}.
Secondary experimental consequences of IVC order have also been proposed which involve orbital magnetization that is modulated on the (graphene) atomic scale~\cite{bultinck2020}. 
However, currently, there are no direct probes of local magnetization with atomic resolution. 
Moreover, similar secondary signatures may be expected from atomic scale valley polarization since each valley of graphene has a characteristic orbital magnetization~\cite{xiao2007}. 
To quantify the extent of spontaneous mixing of valleys, I turn to the pentagonal defects in graphene and its derivatives - where electrons in one valley are scattered into the other by the geometry of the defect (see Fig.~\ref{fig}). 
Essentially, the curvature of the 2D material mixes valley eigenstates and explicitly breaks the same U(1) symmetry that IVC order spontaneously breaks. The pentagonal disclinations harbor a quantized flux of the gauge field that identifies the IVC order parameter, just as a current loop harbours a flux of the magnetic field that identifies magnetization.

Using this observation, I identify the conical defects of graphene capped by a pentagonal disclination as a robust direct tunneling probe of IVC order. 
Such defects are integral to the formation of fullerenes, being responsible for the curvature of these so-called bucky-balls. 
Specifically, I consider large fullerenes derived from uniform icosahedra by decorating the 20 faces with graphene sheets and the 12 vertices with pentagonal disclinations. 

As shown in Fig.~\ref{fig:geometry}, the fullerene is suspended from a metallic lead and is held at a voltage bias relative to the flat twisted bilayer, in a conventional source-drain configuration~\cite{wiesendanger1994}. For a small fullerene in the limit of large barrier potential, differential tunneling conductance probes the local (or locally-averaged) density of states. Larger fullerenes with less curvature and smaller barrier potential allow an effective contact-area that is comparable to the inverse of the moir\'e Brillouin Zone area, closer to  the planar tunneling regime where the theoretical ideas of this work are grounded. Intuitively, I imagine the curved surface of a large buckyball pressing against a flat sample, slightly bending the sample and slightly flattening the fullerene to achieve sufficient contact area. This is to be distinguished from the point-contact spectroscopy limit of an atomically-sharp tip with small barrier potential. A large number of engineering challenges relating to design feasibility are acknowledged and deliberately omitted from the conceptual demonstration of pentagonal defects as a tunneling probe of IVC order in this work.  

First, in Section \ref{secHexTunnel}, I describe the model for planar tunneling from MATBG sample and fullerene tip, and show how  the experimentally observed tunneling current is sensitive to the change in the eigenfunctions of the sample that define the onset of IVC order, through the tunneling matrix element. 
Next, in Section \ref{secPentaTunnel}, I calculate the matrix elements in two tunneling configurations to demonstrate their sensitivity to IVC on very general grounds, independent of details of the modeling. In one configuration, the matrix element decreases with the onset of IVC, in the other configuration it increases by a similar factor. Finally, in Section \ref{secDisc}, I summarize with a discussion on the possibilities that lie at the interface of scanning tunneling spectroscopy and planar tunneling spectroscopy through disclinations.

\section{Continuum model for planar tunneling}\label{secHexTunnel}

I consider a continuum model of the bandstructure of the MATBG sample~\cite{lopesdossantos2007,bistritzer2011,lopesdossantos2012} 
\begin{align}
    H_{\rm TBG}&=-iv^\star\partial_{x}\sigma_1 -iv^\star\partial_{y}\sigma_y\chi_3 + V \chi_2
    \label{eqtbg}
\end{align}
where $\sigma_i$ and $\chi_i$ are the Pauli matrices corresponding to the graphene sublattice and valley degrees of freedom respectively and $V$ is the order parameter for inter-valley coherence. 
This is identical to the banstructure of graphene, except for a renormalized velocity $v^\star/v=\frac{1-3\alpha^2}{1+6\alpha^2}$, where $v$ is the Dirac velocity of monolayer graphene.

Next, I consider the continuum description of the fullerene ``tip'' when its hexagonal facet points toward the sample, as in Fig~\ref{fig}(c). In the limit of a large enough flat facet, the continuum model of graphene is a faithful representation of the ``tip''
\begin{align}
    H^{\rm hex}&=-iv \partial_{x}\sigma_1 -iv \partial_{y}\sigma_2\chi_3
\end{align}

Finally, I consider a continuum description of the fullerene ``tip'' when its pentagonal disclination points towards the sample, as in Fig.\ref{fig}(d). At the simplest level, such a conical defect is modeled by the introduction of two gauge potential vortices~\cite{gonzalez1992, gonzalez1993, vozmediano2010} in the continuum model of graphene
\begin{align}
    -i\nabla \to -i\nabla -  \gamma {\bf A}({\bf r})
\end{align}
where $\gamma$ is a 4$\times$4 matrix in the space of graphene sublattice and valley degrees of freedom, that captures the effect of the gauge potential ${\bf A}({\bf r})$, which I describe by a Aharonov-Bohm flux tube at the origin
\begin{align}
    {\bf A}({\bf r}) = \frac{\Phi}{2\pi} \frac{(-y,x)}{x^2+y^2}
\end{align}
where $\Phi$ represents the strength of the vortex. 
The parameters $\gamma,\Phi$ are obtained by comparing the circulation of the vector potential $\oint {\bf A}\cdot\rm{d}{\bf r}$ with the non-Abelian phase obtained by parallel transporting the 4-component spinor around the defect.

We can identify two such phases, $\exp(\frac{2\pi i }{6} \sigma_3)$ coming from a frustration of the A-B sublattice as we go around the defect, and another $i\chi_2$ coming from the change of the valley index as we circumnavigate the disclination~\cite{gonzalez1992, gonzalez1993, vozmediano2010}. 
Incorporating the corresponding gauge potentials results in the Hamiltonian
\begin{align}
    H^{\rm pent}=H^{\rm hex}+v\left(\frac{5}{6}\sigma_3+\frac{1}{4}\chi_2\right) \frac{-y\sigma_1+x\sigma_2\chi_3}{x^2+y^2}
\end{align}
which in radial coordinates simplifies to
\begin{align}
    -\frac{H^{\rm pent}}{v}=i \sigma_1 \mathcal{U}_\theta \partial_r
      +\frac{i\sigma_2\chi_3}{r} \left(
      \mathcal{U}_\theta^\dagger \partial_\theta + i\mathcal{U}_\theta^\dagger \frac{5}{6}\sigma_3 + i\mathcal{U}_\theta \frac{1}{4}\chi_2\right)\label{eq:7}
\end{align}
where $\mathcal{U}_\theta=\exp(i\theta\sigma_3\chi_3)$.

Next, I assume that the conical defect is flattened by contact with the MATBG sample, to the extent that in-plane momentum is well-defined for most of the area of contact, except at the defect center. This allows us to draw intuition from planar tunneling, while recognizing that the curvature of the fullerene and the spread of the vortex field $\nabla\times{\bf A}_q({\bf r})$ are hard problems that we do not attempt to model with any analytical ansatz.

The full Hamiltonian
\begin{align}
    H=H_{\rm TBG}+H_{\rm tip}+\Delta U(\mathrm{r})
\end{align}
has four cases of interest: IVC order parameter $V=0$ or $V\neq0$ in $H_{\rm TBG}$, and $H_{\rm tip}=H^{\rm hex}$ and $H_{\rm tip}=H^{\rm pent}$ corresponding to the two tunneling configurations represented by the insets of Fig.\ref{fig}(c,d). Here the barrier potential $\Delta U (\bf{r})$ is assumed to be a small perturbation to $H_{\rm TBG}$ and $H_{\rm tip}$. $\Delta U (\bf{r})$ is further assumed to be independent of in-plane coordinate over a large contact area comparable to the inverse size of the Moir\'e Brillouin zone. These are the assumptions under which any intuition from the conceptual anchor-point of planar tunneling is reliable.

The strategy of this demonstration is to use Fermi's golden rule to model the tunneling rate across the  barrier potential $\Delta U (\bf{r})$. Key to this tunneling rate, is the matrix element
\begin{align}
    M(\mathbf{k}_\parallel)=\int_z \langle \psi^{\rm TBG}_{\mathbf{k}_\parallel} 
    \mid
    \Delta U (z)
    \mid 
    \phi^{\rm tip}_{\mathbf{k}_\parallel} \rangle 
\end{align}
between the tip and the sample eigenstates, about which we can make general statements independent of the details of the experimental geometry and its numerical modeling. If the density of states (DOS) across the transition is known from other measurements, the tunneling current normalized by the tip and sample DOS, indicated in the insets of Fig.\ref{fig}(c,d) is proportional to this matrix element $M(\mathbf{k}_\parallel)$, which directly probes the change in the Bloch wavefunctions at the onset of IVC ordering. 

\section{Matrix elements of tunneling amplitude}\label{secPentaTunnel}

In the absence of IVC, it is clear that $H_{\rm TBG}(V=0)$ in \eqref{eqtbg} has the same degenerate eigenstates as monolayer graphene with  energy $E=v^\star k$~\cite{lopesdossantos2007,bistritzer2011,lopesdossantos2012} and eigenfunctions 
\begin{align}
    \psi_1(\mathbf{k})=\frac{1}{\sqrt{2}}
    \begin{pmatrix}
    z\\
    z^{-1}\\
    0 \\
    0
    \end{pmatrix},
    \psi_2(\mathbf{k})=\frac{1}{\sqrt{2}}
    \begin{pmatrix}
    0\\
    0\\
    z^{-1}\\
    z
    \end{pmatrix}    
\end{align}
in the basis 
$| {\bf K_M+k},A \rangle$, 
$| {\bf K_M+k},B \rangle$, 
$| {\bf K'_M+k},A \rangle$, 
$| {\bf K'_M+k},B \rangle$ 
where ${\bf k}$ are crystal momenta in the moir\'e Brillouin zone with corners at ${\bf K_M,K'_M}$, where $z=e^{i\phi},\phi=\tan^{-1}(k_y/k_x),k=\sqrt{k_x^2+k_y^2}$. 
In the symmetry-broken state, degenerate perturbation theory in $V/(v^* k)$ provides the perturbed eigenstates of $H_{\rm TBG}(V \neq 0)$
\begin{align}
    \psi_\pm({\bf k}) = \frac{\psi_1({\bf k}) \mp i \psi_2({\bf k})}{\sqrt{2}},\,\, E_\pm ({\bf k}) = v^* k \pm V.
\end{align}

Next, I recall that the eigenstates of $H^{\rm hex}$ are 
\begin{align}
    \phi_1(\mathbf{k})=\frac{1}{\sqrt{2}}
    \begin{pmatrix}
    z\\
    z^{-1}\\
    0 \\
    0
    \end{pmatrix},
    \phi_2(\mathbf{k})=\frac{1}{\sqrt{2}}
    \begin{pmatrix}
    0\\
    0\\
    z^{-1}\\
    z
    \end{pmatrix}    
\end{align}
in the basis 
$| {\bf K+k},A \rangle$, 
$| {\bf K+k},B \rangle$, 
$| {\bf K'+k},A \rangle$, 
$| {\bf K'+k},B \rangle$ with energy $E(\mathbf{k})=v k$.

The overlap between these wavefunctions determines the experimentally measurable matrix element $M({\bf k})=\int_z \Delta U (z) \langle \psi_{\bf k}^{\rm TBG} | \phi_{\bf k}^{\rm tip}\rangle$. When $H_{\rm tip}=H^{\rm hex}$, and the alignment is such that ${\bf K}\approx{\bf K_M}$, we can identify two distinct behaviours of the matrix element
\begin{align}
    \langle \psi_1 | \phi_1 \rangle &=1,\; &{\rm if}\;V=0\notag\\
    \langle \psi_+ | \phi_1 \rangle &=\frac{1}{\sqrt{2}},\; &{\rm if}\;V\neq0 \label{eq:pred1}
\end{align}
Note that the eigenstates $\psi_\pm$ are non-degenerate and their overlap with $\phi_1$ is spectroscopically resolvable. 

Finally, I address the nature of the eigenstates of $H^{\rm pent}$. 
The key point is that the presence of the $\chi_2$ term in \eqref{eq:7} means that the vortex-bound states are not valley-charge conserving. 
A generic bound-state wavefunction $\phi({\bf r})$ can be expanded in the basis $\phi_1({\bf  k}),\phi_2({\bf  k})$ and their orthonormal complements $\phi_3({\bf  k})=(z,-z^{-1},0,0)^T/\sqrt{2},\phi_4({\bf  k})=(0,0,z^{-1},-z)^T/\sqrt{2}$ as
\begin{align}
    \phi({\bf r}) = \int_{\bf k} &\alpha_{\bf k}({\bf r}) \phi_1({\bf k}) + \beta_{\bf k}({\bf r}) \phi_2({\bf k}) \notag\\
    &+ \gamma_{\bf k}({\bf r}) \phi_3({\bf k}) + \delta_{\bf k}({\bf r}) \phi_4({\bf k})
\end{align}
with a spread in real space determined by the inverse spread in the form-factors $\alpha_{\bf k}$. These form-factors are determined by solving \eqref{eq:7} with the boundary conditions imposed by the leads at large $\bf{r}$, appropriately modified by the curvature effects discussed earlier, which become important at small $\bf{r}$.

The low-energy extended eigenstates of $H^{\rm pent}$ scattering off these vortex-bound states develop into resonances which mix valley eigenstates
\begin{align}
    |\Tilde{\phi}_{\bf k}\rangle \approx \sum_{\bf G} 
      &\Tilde{\alpha}_G |\phi_{1,{\bf k+G}}\rangle 
    + \Tilde{\beta}_G |\phi_{2,{\bf k+G}}\rangle \notag \\
    &+ \Tilde{\gamma}_G |\phi_{3,{\bf k+G}}\rangle
    + \Tilde{\delta}_G |\phi_{4,{\bf k+G}}\rangle\label{eq:resonance}
\end{align}
where ${\bf G}$ runs over the reciprocal lattice vectors. Intuitively, the scattering process is similar to the motion of an asteroid or spacecraft in the vicinity of a static gravitational field, mimicked by the curvature of the fullerene in the vicinity of the disclination. The incoming object may change direction, analogous to change of valley index, or gain momentum as in a gravitational slingshot or gravity-assist process~\cite{negri2020}, analogous to the Umklapp processes included in \eqref{eq:resonance}. In the weak-scattering limit, we expect dominant forward-scattering, which implies $\widetilde{\alpha}_0\to 1$ while other components vanish. 

The overlap of these states with the eigenstates of $H_{\rm TBG}$ determine the experimentally measurable matrix element $M({\bf k})=\int_z \Delta U (z) \langle \psi_{\bf k}^{\rm TBG} | \phi_{\bf k}^{\rm tip}\rangle$. Under similar assumptions as above, when $H_{\rm tip}=H^{\rm pent}$, we can identify two distinct behaviours
\begin{align}
    \langle \psi_1|\widetilde{\phi} \rangle &=\widetilde{\alpha}_0,\, \langle \psi_2|\widetilde{\phi} \rangle =\widetilde{\beta}_0\; &{\rm if}\; V=0\notag\\
    \langle \psi_+|\widetilde{\phi} \rangle &=\widetilde{\alpha}_0-i\widetilde{\beta}_0,\, \langle \psi_-|\widetilde{\phi} \rangle =\widetilde{\alpha}_0+i\widetilde{\beta}_0\; &{\rm if}\; V\neq0 \label{eq:pred2}
\end{align}
and the observable $|\langle \psi_{\bf k}^{\rm TBG} | \phi_{\bf k}^{\rm tip}\rangle|^2$ now \emph{increases} at the onset of IVC order. Note that the experimentally relevant quantity is the differential conductance \emph{per state}, which is $\sim 0.5\times(|\widetilde{\alpha}_0|^2+|\widetilde{\beta}_0|^2)$ for $V=0$, averaged over the two degenerate states. This signal increases at the onset of IVC in this configuration as much as it decreases in the other configuration.

\section{Discussion}\label{secDisc}

I have demonstrated the opposing tendencies of the tunneling matrix elements at the onset of IVC order for two tunneling configurations. 
For large fullerenes oriented with the flat facets contacting the sample, the differential conductance normalized by the density of states decreases with IVC ordering, as the overlap with Bloch wavefunctions of one valley reduces  as indicated in \eqref{eq:pred1}.
On the other hand, the same signal increases with IVC ordering when the fullerene is oriented such that the corner facets capped by the pentagonal disclinations contact the sample. This is because the overlap with valley-mixed modes near the disclination increases with IVC ordering, as indicated in \eqref{eq:pred2}. The overlap with valley-mixed states is as direct a probe of IVC order as the deflection of a compass needle or spin-selective tunneling is a probe of the onset of finite magnetization. This is to be contrasted with traditional atomically resolved tunneling microscopy, which is sensitive to the local density of states.

There is the possibility that tunneling from a single large fullerene may realize both these opposing tendencies, if it can be rotated so that the hexagonal facets or the pentagonal disclinations face the MATBG sample. 
This allows robust quantifiable detection of the onset of inter-valley coherence in moire materials. 
Moreover, if these buckyballs can be suspended on a scanning tip, they realize a ``local'' probe of valley mixing - provided the contact area is large enough for the valley to be well-defined during the tunneling process.
Fullerenes smaller than the moire lengthscale can allow intra-unit-cell spatial resolution of the IVC order parameter. Creatively attaching leads to certain azimuthal sectors of the sample and tip in this planar tunneling geometry may even allow the possibility of valley-selective scanning tunneling spectroscopy. 

{\bf Acknowledgements:} I am grateful for insightful discussions with Francisco Guinea, Paula Mellado, Kevin Nuckolls and the hospitality of the International Centre for Theoretical Physics, Trieste, Italy where this work was conceived. I was supported by the Alexander von Humboldt Foundation.

\bibliographystyle{apsrev4-1}
%\begin{thebibliography}{99}
\bibliography{disclinations}
%\end{thebibliography}
\end{document}